\documentclass[twocolumn,pre,superscriptaddress,preprintnumbers,amsmath,amssymb,longbibliography]{revtex4-1}
\usepackage{graphicx}
\usepackage{color}

\newcommand{\beq}[1]{\begin{equation}\label{#1}}
\newcommand{\eeq}{\end{equation}}
\newcommand{\beqa}[1]{\begin{eqnarray}\label{#1}}
\newcommand{\eeqa}{\end{eqnarray}}

\renewcommand{\rho}{\varrho}
\renewcommand{\epsilon}{\varepsilon}

\def \aap{Astron\&Astroph }
\def \apj{Astrophys J }
\def \apjl{Astrophys J Lett }

\def \mnras {MNRAS }
\def \prl{Phys Rev Lett }
\def \ssr{Space Science Rev.}
\def \araa{Annual Review of Astron and Astrophys}

\begin{document}

\title{Nonaxisymmetric Hall instability: A key to understanding magnetars}

\author{ K. N. Gourgouliatos}
\affiliation{Department of Physics, University of Patras, Patras 26500, Greece;\\
Department of Mathematical Sciences, Durham University, Durham DH1 3LE, UK}
\author{Jos\'e A. Pons}
\affiliation{Departament de F\'{\i}sica Aplicada, Universitat d'Alacant, Ap. Correus 99, 03080 Alacant, Spain }

\date{\today}

\begin{abstract} 

It is generally accepted that the non-linear, dynamical evolution of magnetic fields in the interior of neutron stars plays a key role in the explanation of the observed phenomenology. Understanding the transfer of energy between toroidal and poloidal components, or between different scales, is of particular relevance. In this letter, we present the first 3D simulations of the Hall instability in a neutron star crust, confirming its existence for typical magnetar conditions. We confront our results to estimates obtained by a linear perturbation analysis, which discards any interpretation as numerical instabilities and confirms its physical origin. Interestingly, the  Hall instability creates locally strong magnetic structures that occasionally can make the crust yield to the magnetic stresses and generates coronal loops, similarly as solar coronal loops find their way out through the photosphere. This supports the viability of the mechanism, which has been proposed to explain magnetar outbursts.  
\end{abstract}


\maketitle

Among the many observational faces of neutron stars, {\it magnetars} --  highly magnetized, slowly rotating, isolated neutron stars that sporadically show violent transient activity -- have received much attention in the last decade. Understanding their puzzling behaviour, arguably caused by their magnetic activity, is one of the most active research areas of neutron star physics. In the magnetar scenario, one usually appeals to the creation or existence of small scale magnetic structures to justify the observed phenomenology \cite{Rea:2011,perna11,mereghetti15,coti18}. For instance, observational evidence \cite{tiengo13} favors small structures emerging from the surface, similar to the sunspots in the solar corona \cite{beloborodov07}.
The details about how exactly such small-scale magnetic structures are created are under debate,  but 
they must have their origin in the dynamics of the interior, in particular of the neutron star crust.

The evolution of the magnetic field in a neutron star (NS) crust is governed by the combined action of Ohmic dissipation and the Hall drift \cite{goldreich92,cumming04}. In the presence of a strong field, as in magnetars, the non-linear Hall drift dominates and has been proposed to be the main responsible for the generation of small scales through the so-called {\it Hall instability} 
\cite{RG2002} (hereafter RG02). The occurrence of the Hall instability in a real neutron star has been somewhat controversial, in part because of the lack of numerical simulations able to reproduce the strong-field regime under realistic conditions. The first 2D simulations of the magnetic field evolution in NS crusts \cite{Hollerbach2000,HR2002,Pons2007,gourgouliatos14a,gourgouliatos14b} 
did not find strong evidence of such instability, but they were restricted to axisymmetric models and moderate magnetization, for numerical reasons. 
A first 3D non-linear study in a periodic box \cite{wareing2009} concluded that any instabilities are overwhelmed by a turbulent Hall cascade. Conversely, \cite{pons2010} confirmed the occurrence of the instability showing that, because the unstable modes have a relatively long wavelength, it is suppressed on a cubic domain with periodic boundary conditions, but it 
arises on a thin slab where one of the spatial lengths is longer than others. And this is precisely the geometry of a neutron star crust, a thin spherical shell ($\approx 1$ km) 
of radius $R\approx 12$ km.

We must note the distinction between the {\it resistive} Hall instability, which is essentially a tearing mode \cite{Fruchtman:1993,gourgouliatos16b}, and {\it ideal} instabilities which operate under infinite conductivity, for example, the density-shear instability requiring a density gradient \cite{Wood:2014, gourgouliatos15b}, or the fast collisionless reconnection observed  in the whistler frequency range  \cite{Attico2000}.
Indeed, the growth times of the Hall instability modes become increasingly large with vanishing resistivity.
Another relevant issue is that, in spherical geometry, the instability may be suppressed for the axisymmetric modes. Thus, 2D simulations could not help 
resolving the controversy, and only recently \cite{wood2015,gourgouliatos16,vigano19}, 3D simulations have been possible.

In this letter, we present the first 3D simulations of the Hall instability for a model with a strong toroidal field in a NS crust, confirming its occurrence. We show how small magnetic structures are naturally created and drift toward the star's poles. Although similar structures had been observed before in previous simulations \cite{wood2015, gourgouliatos16, gourgouliatos18}, their actual  origin and the true nature of a possible instability had not been settled.  
We further present a linear stability analysis in a spherical shell that gives similar results to the non-linear simulations, thus 
reinforcing our conclusion that the Hall instability is actually at the origin of the observed magnetar activity.

To a very good approximation, the crust can be considered a one component plasma, where only electrons can flow through the ionic lattice. Since ions are fixed, there is no mass 
motion, and the induction equation reduces to:
\begin{equation} 
\frac{\partial\vec B}{\partial t}= - {\nabla \times}\left(\eta 
\nabla \times \vec{B}  + \frac{c}{4 \pi e n_e} (\nabla \times \vec{B} ) \times \vec B  \right) ,
\label{Hallind} 
\end{equation} 
where $c$ is the speed of light, $\eta=c^2/4\pi \sigma$ is the magnetic diffusivity, $\sigma$ is the electrical conductivity, $e$ is the elementary charge and $n_e$ is the electron number density.
We consider a background toroidal field of the form
\begin{eqnarray}
B_0 &=& \sin\theta \frac{ f(r)}{r} \vec{e}_\phi  ~,
\label{B0}
\end{eqnarray}
and evolve it by numerically solving eq.~(\ref{Hallind}) inside a spherical shell, using a version of the PARODY 3-D MHD code \cite{Dormy:1998, Aubert:2008}
suitably adapted to NSs  \cite{wood2015,gourgouliatos16}. We impose vacuum boundary conditions at the exterior of the star and superconductor boundary conditions at the base of the crust, not allowing the magnetic field to penetrate into the core. We consider a crust of uniform electron number density $n_e=2.5\times 10^{34}$cm$^{-3}$, electric conductivity $\sigma=1.8\times 10^{23}$s$^{-1}$, and we express the magnetic field in units of $B_{14}=B/10^{14}$G. 
We have decided to study simplified models with constant density and conductivity for two reasons. First, this allows us to distinguish between the families of resistive and ideal instabilities that may operate in the electron-MHD regime. Namely, a realistic crust with stratified density and composition is subject to both instabilities, whereas our set up suppresses the ideal density-shear instability \cite{gourgouliatos15b}, but permits the resistive one \cite{RG2002} to grow. Second, this choice also allows us a more direct comparison to previous works.

\begin{figure}
\includegraphics[width=0.45\textwidth]{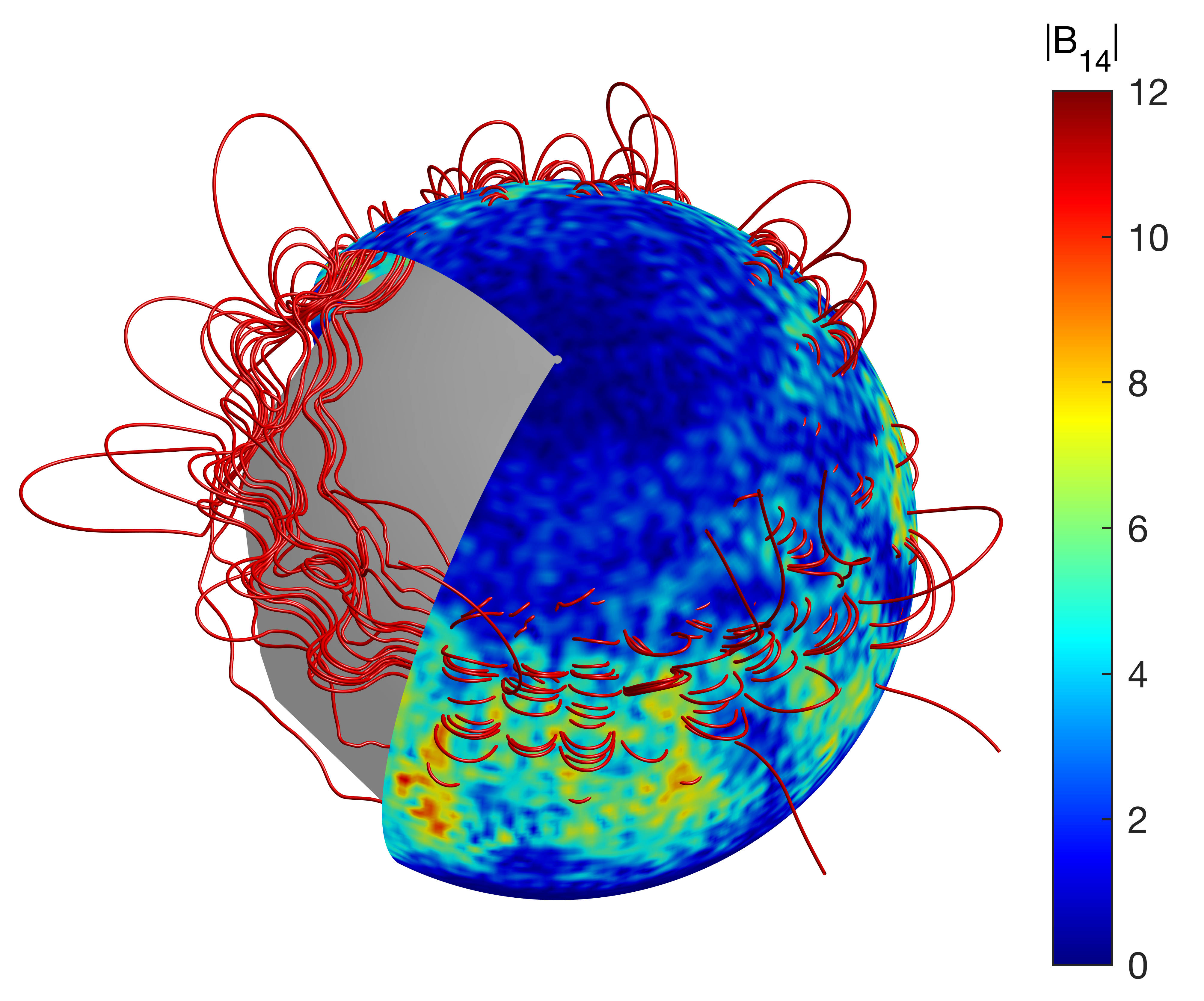}
\caption{Magnetic field lines and magnetic field intensity map (in color scale) on the star surface, at $t=3$ kyrs for model A.}
 \label{Fig1}
\end{figure}

To explore how our results depend on the thickness of the layer where the magnetic field is confined, 
we have considered two cases: a realistic one where the inner radius of the crust $R_{in}=0.9 R_{NS}$ (hereafter model A) and another with $R_{in}=0.5 R_{NS}$ (hereafter model B). 
For structures with a length-scale $L$, measured in km, the Hall timescale is $t_H= 4 \pi e n_e L^2/cB =16 L^{2} / B_{14}$ kyr, and the Ohmic dissipation timescale $t_{D} = 4 \pi \sigma L^2/c^2 = 0.8 L^2$ Myr. The ratio $t_D/t_H$ (equivalent to the magnetic Reynolds number) is $50 B_{14}$.

In general, purely azimuthal magnetic fields, as our background field (\ref{B0}), cannot be in Hall equilibrium; however, their evolution maintains their azimuthal structure and does not generate any radial or meridional component, a result that has been confirmed by axisymmetric simulations. This is also the case in 3-D simulations provided that non-axisymmetric perturbations are not included in the initial conditions. Therefore, a strong indication of the operation of the resistive Hall instability is the growth of non-axisymmetric structures, even if the background field drifts at the same time in the meridional and radial direction. 

\begin{figure}
\includegraphics[width=0.45\textwidth]{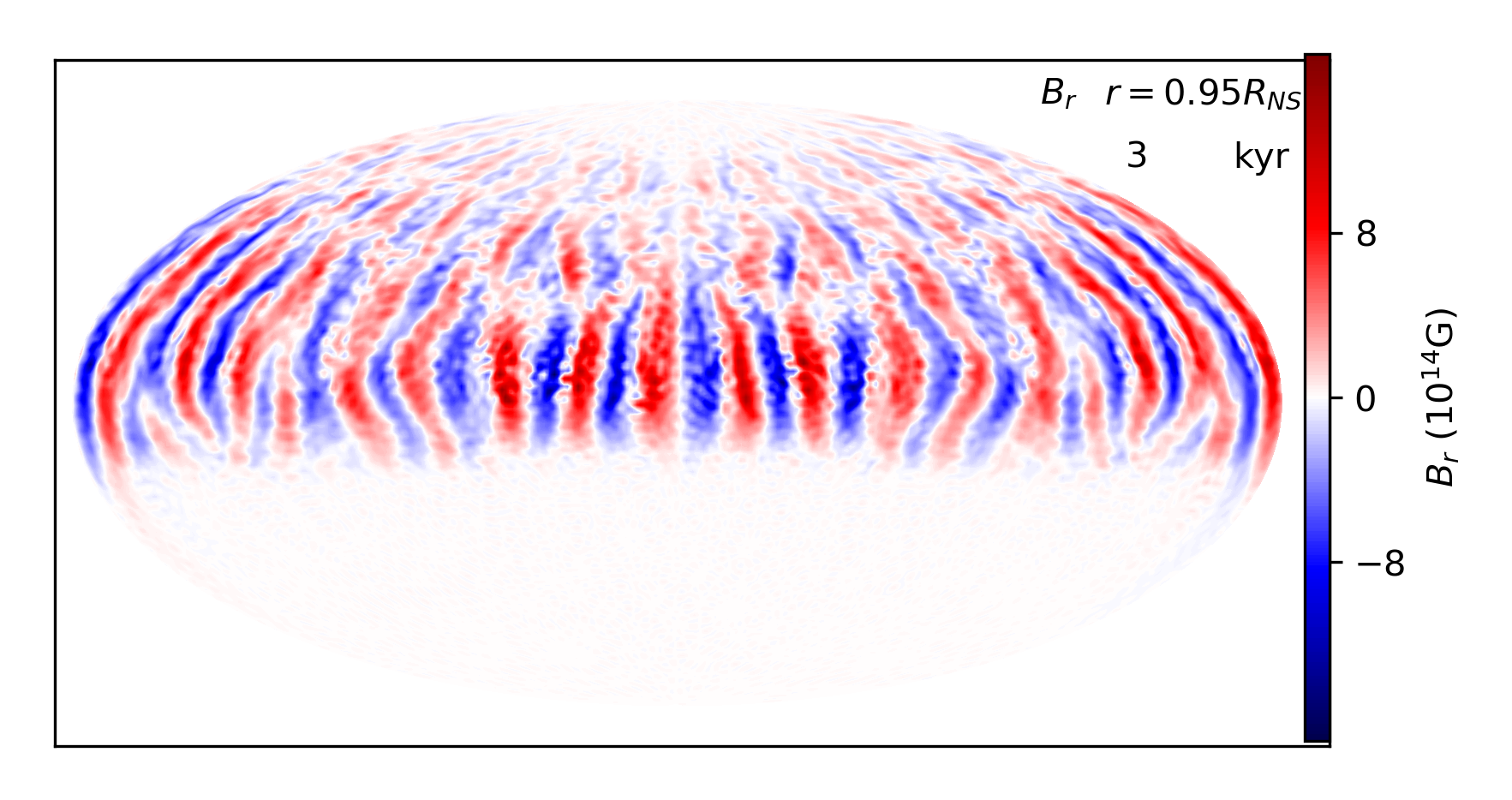}
\includegraphics[width=0.45\textwidth]{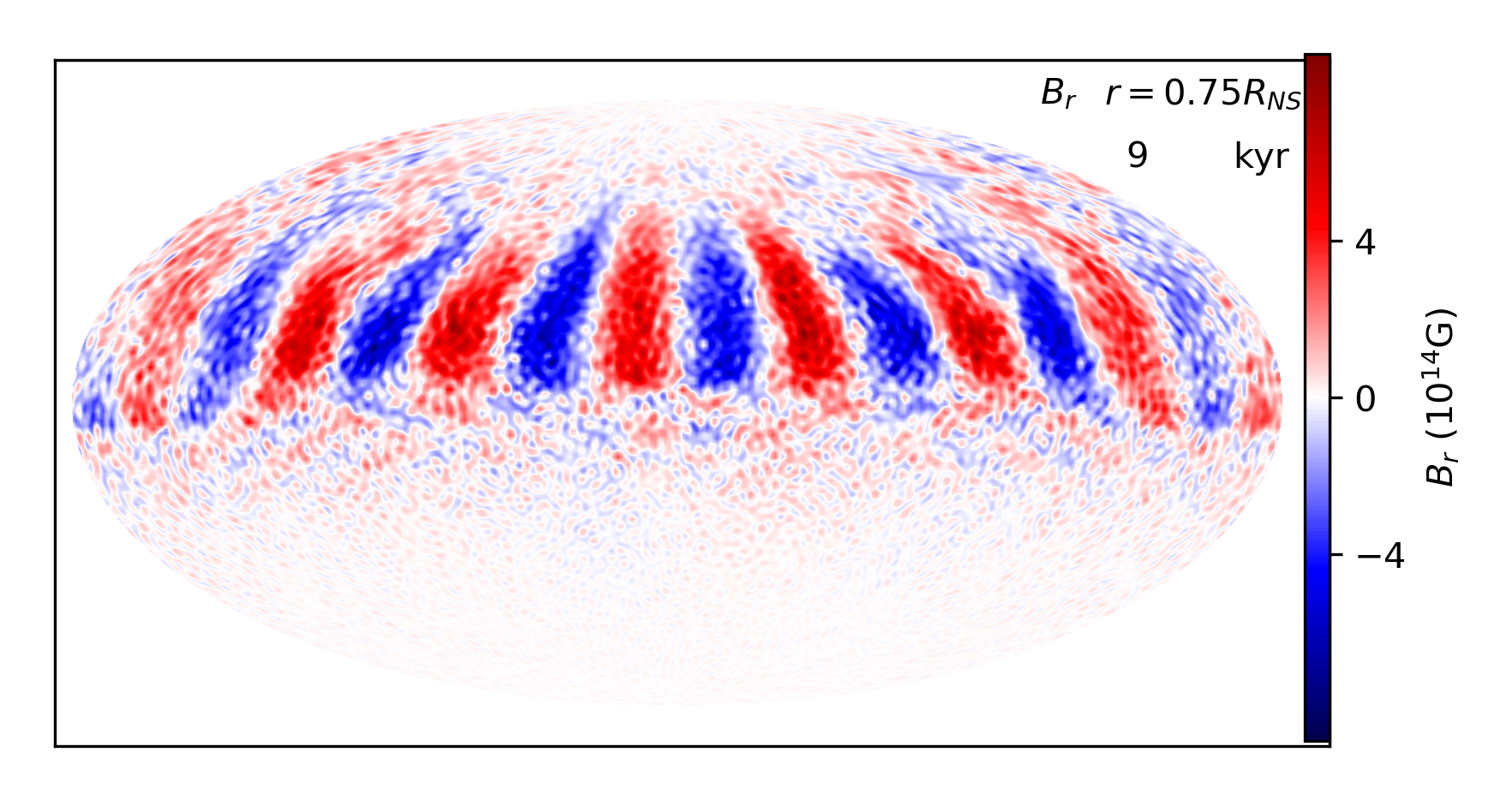}  
\caption{The radial component of the magnetic field for models A and B (top and bottom panel) in the middle of the crust ($r=0.95R_{NS}$ and $r=0.75R_{NS}$) at 3 and 9 kyr respectively. }
 \label{Fig2}
\end{figure}

To test this hypothesis, we have performed 3-D simulations of the initial field given by eq.~(\ref{B0}) with 
$$f(r)=\tilde{B}R_{NS}\left(\frac{0.5}{R_{NS}-R_{in}}\right)^3\left(R_{NS}-r\right)^2\left(r-R_{in}\right),$$ 
where $\tilde{B}$ is a normalisation parameter, chosen so that the maximum value attained by the magnetic field inside the crust is $2\times 10^{15}$G. We have used this particular profile due to its similarity to the profile studied in RG02. Furthermore, we also include non-axisymmetric perturbations, with a flat spectrum, exciting the azimuthal $l=m$ modes from $1$ to $40$. 
Starting from these initial conditions, we find that there is a continuous growth of the non-axisymmetric modes with the appearance of zones with alternating inwards and outwards magnetic field. While this is happening, the magnetic field drifts towards the northern hemisphere, as in the axisymmetric simulations. 

\begin{figure*}
\includegraphics[width=0.46\textwidth]{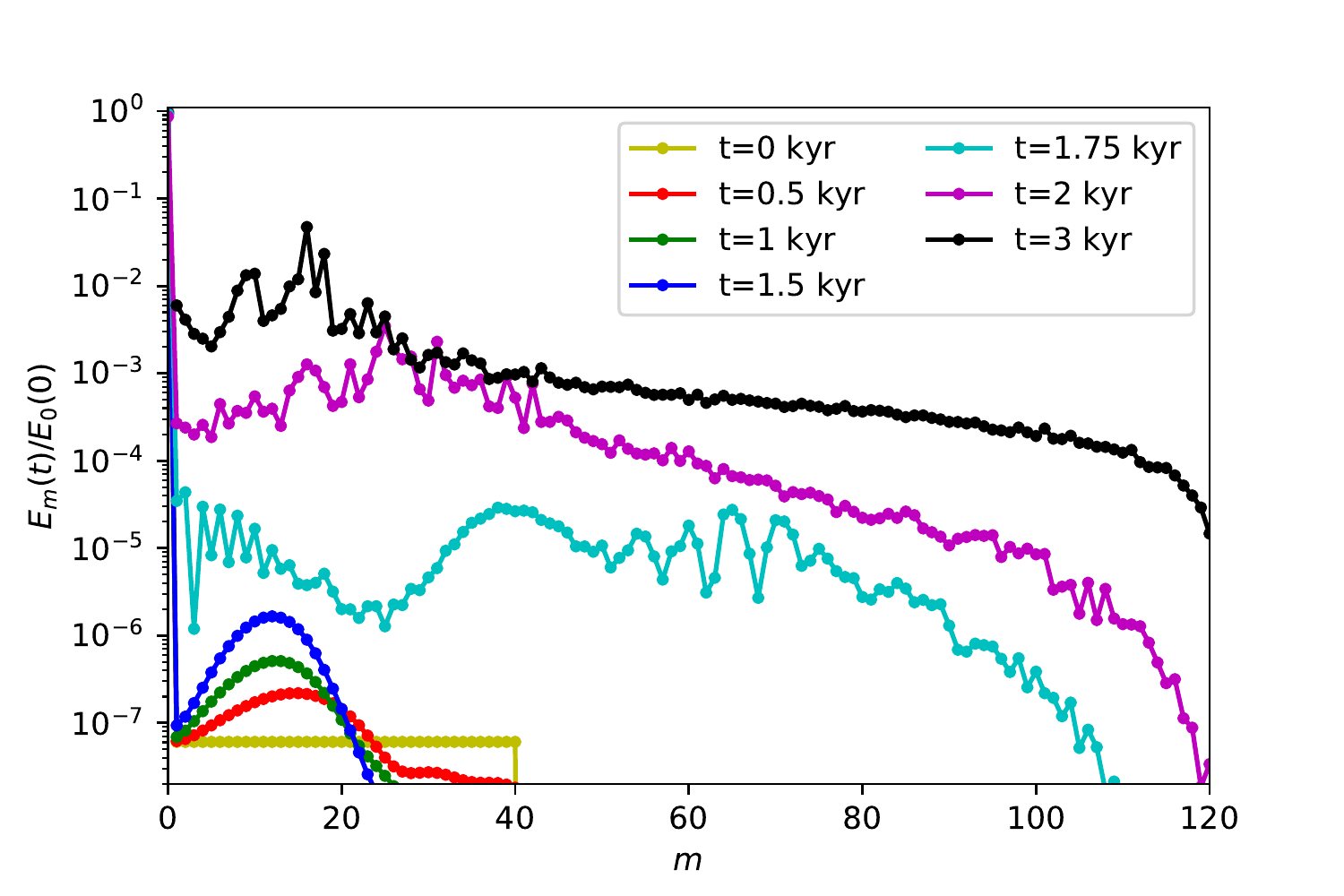}
\includegraphics[width=0.46\textwidth]{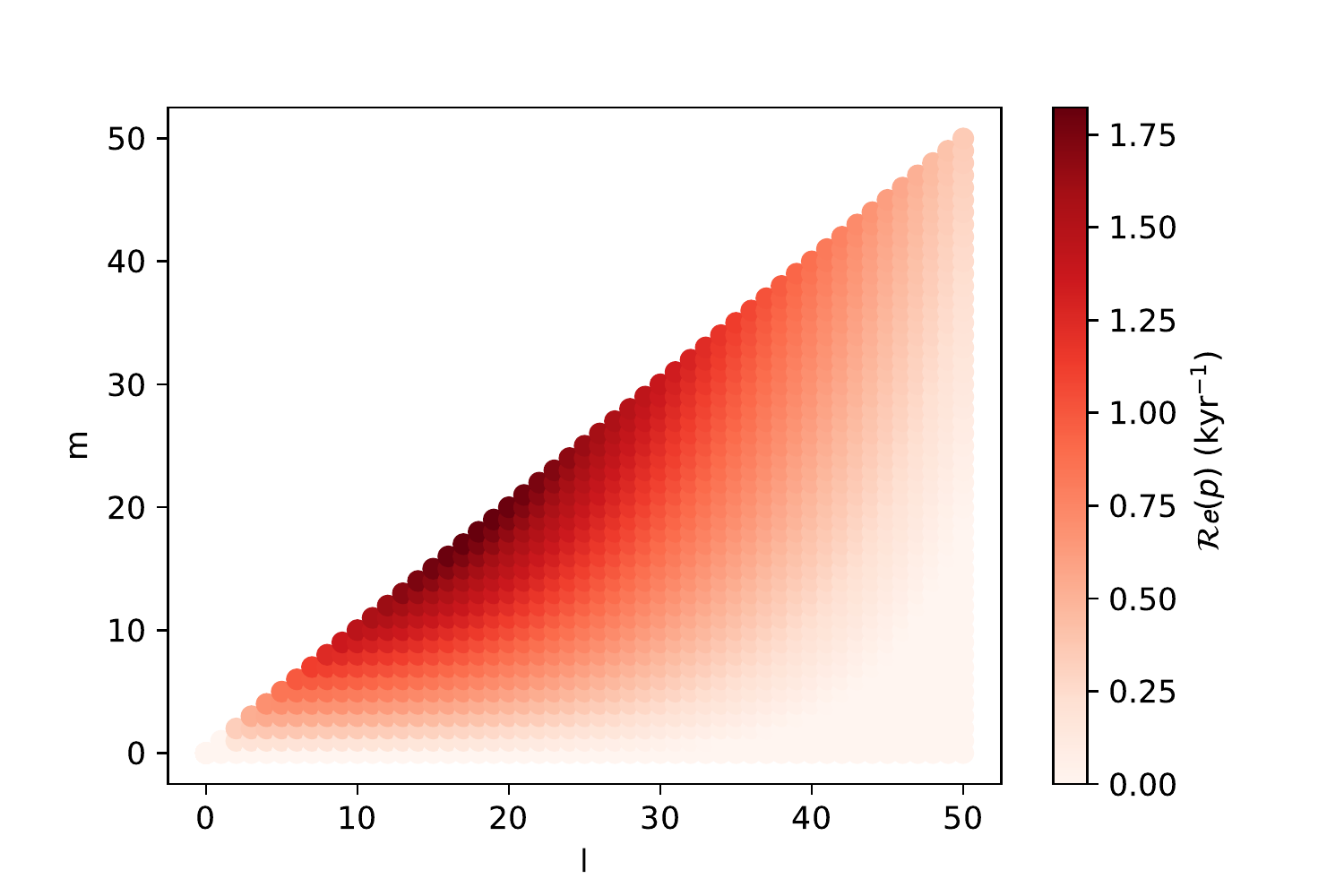}
\caption{Left panel: Spectral power distribution as a function of $m$ (summed over all $l$), obtained from the 3D simulations. Right panel: results from the linear analysis. The color scale shows the inverse growth rates for each mode. }
\label{Fig3}
\end{figure*}

In Fig.~\ref{Fig1} we show an illustration of a snapshot during the evolution of model A. Besides the global drift toward the north pole, we find that many small-scale structures arise. Locally, the magnetic field intensity in some regions is about one order of magnitude higher than the average value. This figure corresponds to model A, but the qualitative behaviour is similar in both models. However, there is a clear quantitative distinction. In model A, the number of structures in the azimuthal direction is $18$, whereas in  model B the number of zones is $8$
as we can see in Fig.~\ref{Fig2}, where we compare their internal structure (at a radius of $r=0.95R_{NS}$ and $r=0.75R_{NS}$ for models A and B respectively), after the saturation of the instabilities.
Besides,  the growth time is consistently slower for model B.
This pattern is visible in spectral space, as the peak of the cumulative distribution is $m \sim 20$ for the thin crust (Fig.~\ref{Fig3}) and $m \sim 10$ for the thicker one.

To confirm that the results of non-linear simulations actually correspond to the so-called Hall-instability, we have performed a linear perturbation analysis
of a background toroidal field of the same form as eq.~(\ref{B0}), in a spherical shell. We decompose the perturbation $\vec{b}$ into its poloidal and toroidal components $\vec{b}=\vec{b}_p+\vec{b}_t$, which in turn can be expressed in terms of two scalar functions $S, T$ as follows:
\begin{eqnarray}
\vec{b}_p = - \nabla \times \left( \vec{e}_r \times \nabla S \right), \quad \quad \vec{b}_t = - \left( \vec{e}_r \times \nabla T \right),
\end{eqnarray} 
where $\vec{e}_r$ is the unit vector in the radial direction. Following the standard procedure, but in spherical coordinates, we
expand  the scalar functions in spherical harmonics:
\begin{eqnarray}
S &=&  e^{p \tau}  \sum_{lm} s_{lm}(r)  Y_l^m(\theta,\phi)  \\
T &=& e^{p \tau} \sum_{lm} t_{lm}(r)  Y_l^m  (\theta,\phi), 
\end{eqnarray}
where we use $\tau$ to denote the time variable to avoid confusion with the toroidal radial function $t(r)$. This notation allows us to directly compare with RG02. Here, $p$ is a complex number to be determined.  The eigenvalue $p$ with the largest positive real part represents the fastest growing mode (with growth time $1/{\mathcal Re}(p)$). With help of an algebraic manipulator, we obtain the following equations for the perturbations:
\begin{widetext}
\begin{eqnarray}
{r^2(ps_{lm} - s_{lm}'') + {l(l+1)} s_{lm}} +  i{m \left(1-\frac{2}{l(l+1)} \right) {f t_{lm}}} + {\cal C}_{l\pm 1}^m &=& 0,
 \nonumber \\
{r^2(p t_{lm} - t_{lm}'') + l(l+1) t_{lm} }  +  i~{m} \left( {f'' s_{lm} - f s_{lm}''  + \frac{l(l+1)}{r^2} ~f s_{lm} }   \right. && \nonumber \\
+ \left. \frac{2}{l(l+1)} \left( f s_{lm}''+ s_{lm}' f' - 2\frac{f s_{lm}'}{r} \right) - 2\frac{s_{lm} f'}{r} \right) + {\cal D}_{l\pm 1}^m &=& 0,
\end{eqnarray}
\end{widetext}
where the primes denote radial derivatives and  $ {\cal C}_{l\pm 1}^m$ and ${\cal D}_{l\pm 1}^m$ are a shorthand for the coupling terms with the $l\pm1$ coefficients 
(always with the same $m$ index). The $ {\cal C}_{l\pm 1}^m$ term only involves combinations of $f s_{l\pm1,m}'$ and $f' s_{l\pm1,m}$, while the ${\cal D}_{l\pm 1}^m$ 
only involves the product $f t_{l\pm1, m}$ and its first radial derivative.

For reference, in Cartesian coordinates and planar symmetry, RG02 expanded the perturbation in plane waves and obtained
\begin{eqnarray}
{ps - s'' + k^2~s} + i ({k_x f t} - k_y s f') = 0  ~,\\
{pt - t'' + k^2~t} + i {k_x (f'' s -f s'' + k^2 f s)}= 0~.
\end{eqnarray}
The similitude between  equations is clearly visible by identifying $k^2 \rightarrow l(l+1)/r^2$, $k_x f  \rightarrow m f/r^2$.  
We note that our definition of $f$ in the spherical case differs from the planar case (RG02) in a factor $r$ ($f$ has units of magnetic field times length).

In principle, one should solve the fully coupled system of equations for a large number of $l,m$'s. However, as a first approximation, and in part motivated by the 
similitude of the equations with the cartesian case, we neglect coupling terms. We also note that, in RG02, the fast-growing mode always has $k_y=0$. This
cancels the term proportional to $s f'$, the counterpart to our coupling ${\cal C}_{l\pm 1}^m$. Our purpose is not to perform a complete and detailed linear analysis, 
but rather to asses on the interpretation of our 3D non-linear simulations.

In Fig.~\ref{Fig3} we compare the results of this truncated linear analysis (right) to the non-linear simulations (left). The right panels show, in a red color scale, the inverse growth time of the unstable models for each $l$ and $m$. Interestingly, the linear analysis (even neglecting couplings between modes) agrees very well with the simulations. For each multipole, the fastest growing modes always correspond to $l=m$. With some visual effort translating the two scenarios, one can compare our results in a spherical shell with Fig 1 in RG02 in a rectangular slab. Note that the correct analogy should compare $l \approx |k|=(k_x^2+k_y^2)^{1/2}$ and
$m\approx k_x$. Therefore, the fastest growing modes with $k_y=0$ in RG02 should correspond to $|k|=k_x$. This is $l=m$, as we obtain here. If we look for example at model A, the analytical estimate of the growth time is $1/{\mathcal Re}(p_{\rm max}) \approx 0.6$ kyrs, in excellent agreement with the observed growth in the left panel  during the linear phase, $t=0, 0.5, $ and 1 kyrs. After a few growth times, the background has begun to change, 
the non-linear evolution sets in, and the full spectrum is filled by the Hall cascade (see curves at $t=1.75, 2$ and 3 kyr), but we still find a significant excess power around the $m=20$ region.
We must note that the linear analysis is a first order approximation (because we omit couplings between neighbour modes), and one should not expect an exact identification of a single fast growing mode in the non-linear results. Moreover, there are a few modes (in the $m=10-20$ range) with very similar growth times. 

The linear growth phase in Fig. 3a corresponds to the first 3 snapshots (t=0,0.5,1 kyrs), and the fact that there is a range of
modes (not exactly peaking at m=18) is not due (yet) to a fast evolution of the background, but to the fact that the linear
analysis is approximate (we truncate couplings with neighbour modes to make it simple). So, this numbers can be taken as a good
indication, but we do not claim that the fast growing mode is exactly m=18, with  t=1.75 kyrs. We can simply conclude that the
most unstable modes are in the range m=10-25, with typical growth rates of 1-2 kyrs.
Similar considerations apply to model B.
We have also obtained a few eigenfunctions and checked that they are qualitatively similar to the eigenfunctions of RG02.

Thus, we confirm that the instability observed in the 3D simulations affects approximately the same wavelength range
and grows on the same timescale of the linear analysis estimates. We should stress that the typical wavelength of the most unstable mode is closely correlated to the
thickness of the shell where the toroidal field is confined. In a realistic crust, we expect structures with $m \approx 20$ and a typical size of $2 \pi R/20 \approx 3-4$ km (and proportionally smaller for toroidal rings shifted to higher latitudes). We have considered a purely toroidal field for simplicity of the analysis, 
but we have obtained very similar results in 3D simulations adding an initial poloidal component and a stratified density \cite{gourgouliatos18}, concluding that the instability operating here is the Hall instability (RG02), rather than the ideal one.

 A critical ingredient for the instabilities is the choice of the appropriate boundary conditions. Throughout this work, we have considered a non-permeating boundary condition at the inner crust and a vacuum potential solution in the outer region. A more realistic approach would consider the role of a magnetic field threading through the core, as at magnetar field strengths the assumption of a Type-I superconductor may not hold. In the exterior, a current filled-magnetosphere relaxing to a force-free equilibrium or even dynamically evolving may be more suitable.  

As our main purpose here is to investigate the development of the instability, we have chosen highly axisymmetric initial conditions, where the energy in the non-axisymmetric component is six orders of magnitude less than the axisymmetric part. Because of that, we see the formation of multiple zones. In a realistic configuration, the initial conditions may not be that symmetric, therefore instead of the excitation of a higher multipolar structure, a less ordered magnetic field configuration may develop.

The main implication of our result is that a sufficiently strong toroidal field, as most magnetized NS models assume,  is subject to this non-axisymmetric instability, and will break into small structures (typically 10-20) in the azimuthal direction. Such structures can occasionally make the crust yield to the magnetic stresses  \cite{perna11,beloborodov14,levin16}, leading to the formation of magnetic loops similar to the solar coronal loops. This mechanism is believed to be at the origin of magnetar outbursts.

We also note that the loops created with this mechanism have magnetic field strengths typically one order of magnitude larger than the dipolar large scale field, in line with the observations \cite{tiengo13}.   Besides, our findings also have implications for the quiescent emission. It is has been proposed  \cite{akgun18} that the high temperatures of magnetars are due to the dissipation of currents in a shallow layer when magnetospheric currents return to close the circuit inside the star (see similar arguments in  \cite{carrasco19,karageorgopoulos19}). The creation of small, force-free magnetic spots with the right size (a few km$^2$) is consistent with the typical sizes of the hot emitting spots of magnetars in quiescence \cite{Turolla:2015, Kaspi:2017}. 
 Further works studying the coupled 3D magnetic and thermal evolution of magnetars are needed to understand when, and how often, one of this spots results in a coronal-like flare and locally high temperatures.\\

This work used the DiRAC@Durham facility managed by the Institute for Computational Cosmology on behalf of the STFC DiRAC HPC Facility (www.dirac.ac.uk). The equipment was funded by BEIS capital funding via STFC capital Grants No. ST/P002293/1, No. ST/R002371/1 and No. ST/S002502/1, Durham University and STFC operations Grant No. ST/R000832/1. DiRAC is part of the National e-Infrastructure.

\bibliographystyle{apsrev4-1}
\bibliography{references}

\end{document}